# Predicting Exercise Adherence and Physical Activity in Older Adults Based on Tablet Engagement: A Post-hoc Study

Sumit Mehra *, Jantine van den Helder, Ben J.A. Kröse, Raoul H.H. Engelbert, Peter J.M. Weijs, Bart Visser

Amsterdam University of Applied Sciences

* contact corresponding author: s.mehra@hva.nl

**Abstract.** Sufficient physical activity can prolong the ability of older adults to live independently. Community-based exercise programs can be enhanced by regularly performing exercises at home. To support such a home-based exercise program, a blended intervention was developed that combined the use of a tablet application with a personal coach. The purpose of the current study was to explore to which extent tablet engagement predicted exercise adherence and physical activity. The results show that older adults (n=133; M=71 years of age) that participated 6 months in a randomized controlled trial, performed at average 12 home-based exercised per week and exercised on average 3 days per week, thereby meeting WHO guidelines. They used the tablet app on average 7 times per week. Multiple linear regressions revealed that the use of the app statistically predicted the number of exercises that were performed and the number of exercise days. Physical activity, however, did not increase and also could not be predicted by exercise frequency or app use. We conclude that engagement with a tablet can contribute to sustained exercise behavior.

## 1      Introduction

Ageing is associated with a decline in daily functioning and mobility [1, 2]. Physical activity can delay the onset and slow down the decline associated with ageing. Older adults that exercise on a regular basis, can prevent functional impairments and prolong the ability to live independently [3, 4]. Various community centers across the world offer senior citizens to exercise on a weekly basis in a group under guidance of an instructor [5–7]. Participating once a week in an exercise group, however, is not sufficient for achieving health benefits [8–10]. WHO guidelines prescribe a higher frequency, intensity and duration of physical activity [11]. Due to the limitations group-based programs face, meeting the guidelines is often not possible [12].

Over the past few years various eHealth or mHealth interventions have been developed to increase physical activity in older adults [13–19]. In order to enhance existing community-based exercise programs, a novel blended intervention, VITAMIN, was developed with end-users [20, 21].. The intervention consisted of a personalized home-based exercise program that was supported by a tablet, in combination with a personal coach [22, 23]. The intervention distinguished itself by a) designed to complement existing community-based programs rather than a stand-alone intervention, b) specifically supports home-based exercises, c) uses blended technology as a mode of delivery and d) design that was theoretically based on behavior change techniques. Furthermore, to increase the efficacy of the exercise program, also nutrition counseling was included. A previously conducted randomized controlled trial (RCT) compared the blended home-based exercise program – with or without nutrition counseling – to a control group that only participated weekly in existing community-based exercise programs. The study showed that during the 6-month intervention period the majority (64.5%) of participants adhered to the recommendation to perform at least two times a week home-based exercises [24]. It remains unclear, however, how participants' engagement with the tablet contributed to exercise adherence. The aim of the present study was not to study the effectiveness of the intervention by comparing it to a control group, which was done in the recently published RCT study [24], but to explore to which extent the tablet engagement predicts exercise adherence in older adults by conducting a secondary analysis of the aggregated data of both RCT groups that received a tablet.

## 2 Methods

### 2.1 Technology

A client-server system was developed that consisted of a front-end tablet application for end-users and a back-end web-based dashboard for coaches that guided the participants. The design of tablet application was based on behavior change techniques that are rooted in self-regulation [22]. The components of the app included an interactive module for goal setting, a library containing over 50 instructional videos of home-based exercises, a possibility to compose a personal training schedule, the ability to track progress and to receive feedback from a personal coach. See Figure 1 to Figure 4. A usability study demonstrated that the intended users were able to operate the app in an effective and efficient manner [21].

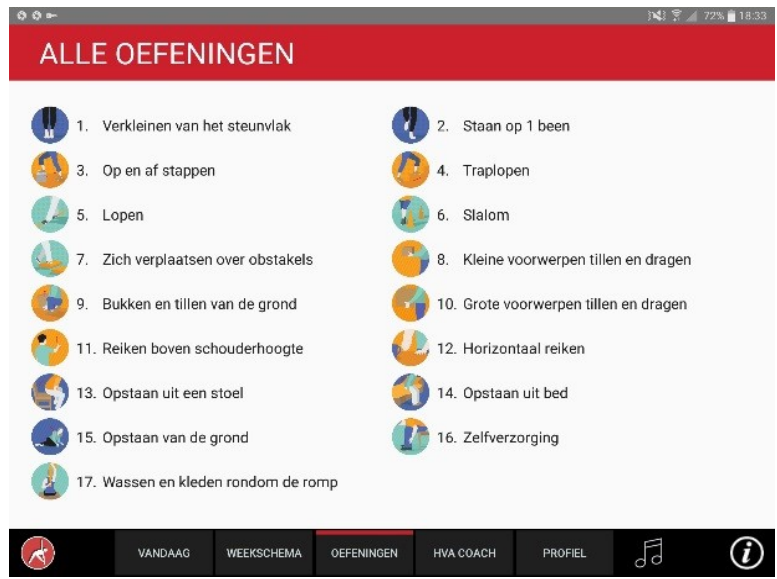

**Fig. 1.** A library of 17 home-based exercises, each available in 3 versions

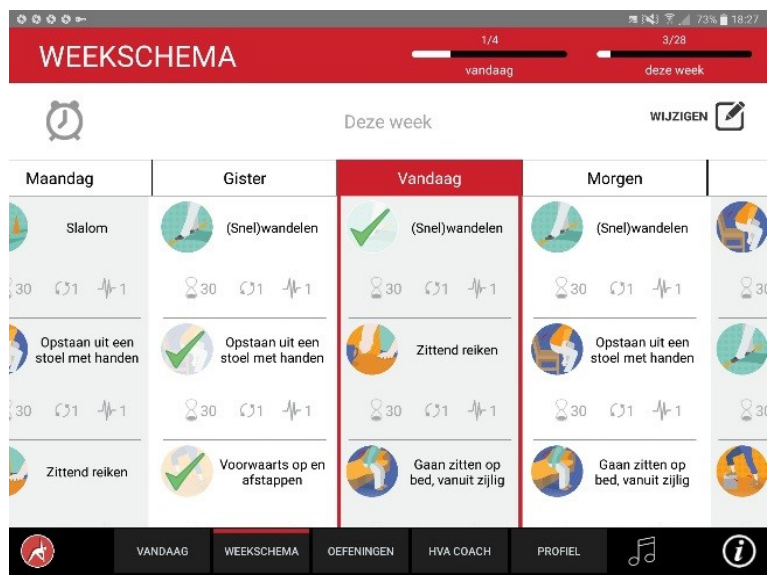

**Fig. 2.** Example of a personalized exercise schedule

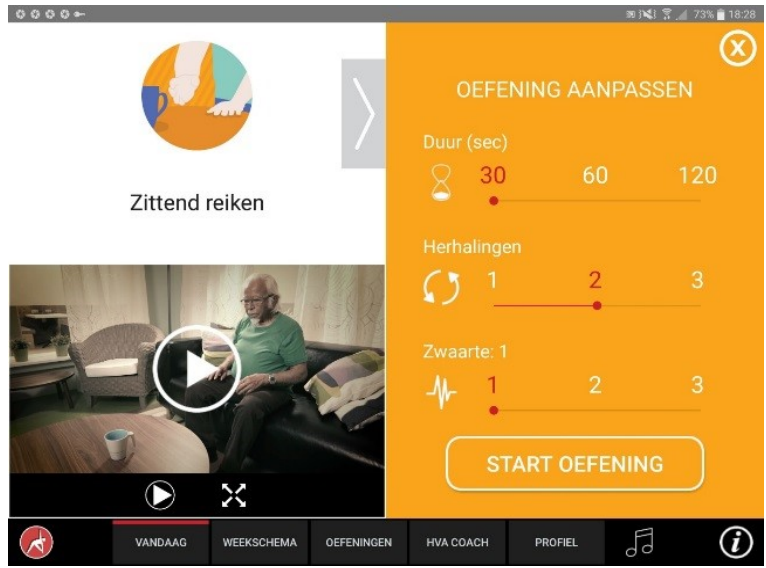

**Fig. 3.** Ability to modify each exercise with three parameters

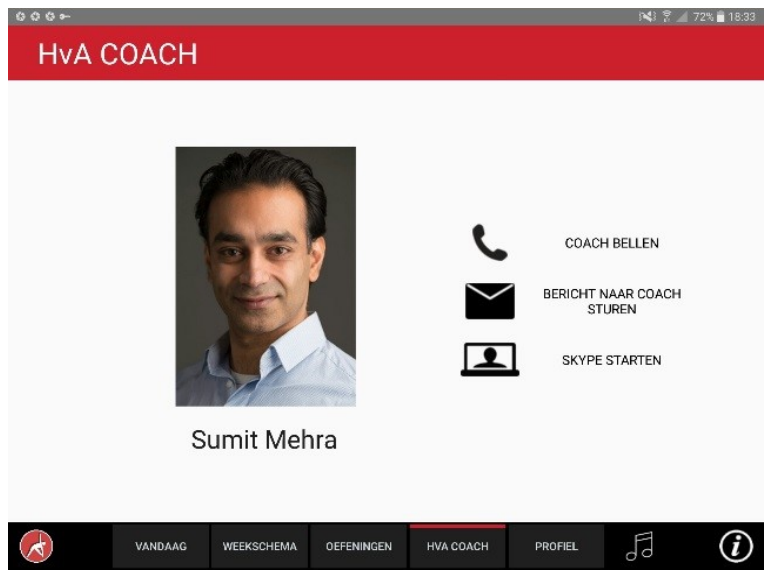

**Fig. 4.** Communication with the coach (telephone call, e-mail, video call)

### 2.2 Study Design and Recruitment

A randomized clinical trial (RCT) was conducted to assess the effectiveness of the blended intervention in terms of health outcomes. It consisted of a 6-month intervention period. The RCT had three arms: 1) a control group of older adults that only participated once a week in group-based exercise classes offered by local community centers, 2) a blended exercise group of older adults that received a tablet and coaching to perform individual home-based exercises, in addition to the weekly group-based exercises and 3) a blended exercise plus nutrition group of older adults that received nutrition counseling, in addition to the weekly group-based exercises and individual home-based exercises. Concerning the home-based exercise program, both intervention groups (arm b and c) were identical. The groups only differed whether or not they received additional nutrition counseling.

Older adults were recruited in the surroundings of Amsterdam, the Netherlands, through either a) addressing visitors of local community centers that were offering weekly group-based exercise classes or b) citizens of Amsterdam through a postal mailing. Applicants were included in the trial if they met the following criteria: 1) 55 years of age or older, which matched the age restrictions the local community centers use, 2) ability to understand the Dutch language and 3) absence of specific cognitive or physical impairments. The protocol that describes the RCT in detail has previously been published [23].

### 2.3 Measures

**Home-based Exercise Adherence.** Participants compiled, with the help of an appointed coach, a personalized exercise program with exercises that varied in the duration, repetitions and difficulty level. Participants were recommended to perform at least two times a week the home-based exercises. During the 6-month intervention period, participants registered with the tablet when they completed their personalized home-based exercises. Based on log data the following frequencies were determined: a) how many days per week they performed exercises, for instance 4 out of 7 days, and b) the total number of exercises they completed per week. To preserve

**Tablet Engagement.** Based on the log data the number of times the tablet app was opened per week was determined (app logins) during the 6-month intervention period.

**Physical Activity Level.** As an exploratory measure, at baseline and after 6 months the participants' physical activity in daily life was assessed by asking participants to keep track of all activities during a period of 3 days using a paper diary. For each participant a physical activity level (PAL) was determined by calculating the average metabolic equivalent of task (MET) per 24 hours. A MET value of 1 represents no physical activity, 1 to 3, 4 to 6 and more than 6 MET represents light, moderate and vigorous intensity activity, respectively.

The difference between exercise and physical activity (PA) is that the term exercises refers to planned, structured, repetitive and intentional movement intended to improve or maintain physical fitness [9, 25]. In contrast, physical activity refers to all activities that involve bodily movement that requires energy. Thus the term physical activity is a broader concept that encompasses exercise, but also includes, for instance, activities like walking, gardening or doing household tasks.

**Motivation.** At baseline and after 6 months the motivation to exercise was measured by the Behavioral Regulation in Exercise Questionnaire-2 (BREQ-2), a validated questionnaire containing 19 Likert items on a 5-point scale [26]. The BREQ-2 distinguishes five forms of motivation derived from the self-determination theory: amotivation 4 items), external regulation (4 items), introjected regulation (3 items), identified regulation (4 items) and intrinsic regulation (4 items).

**Season.** To account for possible seasonal effects, for each participant the offset to midsummer was calculated, by determining the number of days between the date the participant started the intervention and the median of the calendar year (day 183). The minimum score (0) represents midsummer, the maximum score (182) represents midwinter.

**Other.** Age, gender, recruitment strategy and assignment to RCT-arm was recorded for all participants.

**Table 1.** Overview of the measures analyzed during the 6-month intervention period.

| Start of intervention (baseline) | During intervention (0 to 6 months) | End of intervention (6 months) |
|---|---|---|
| Physical Activity Level (PAL) | Exercise adherence:<br>• Number of performed exercises per week<br>• Number of days exercises were performed per week | Physical Activity Level (PAL) |
| Motivation (BREQ-2) | Tablet engagement:<br>• Number of times app has been used per week | Motivation (BREQ-2) |
| Participants & intervention characteristics: age, gender, recruitment strategy, RCT-group, season | | |

**Data analysis**

Data of participants that were assigned to either group 2 (blended exercise program) or group 3 (blended exercise program with nutrition counseling) was aggregated, resulting in a dataset of all trial participants that received a tablet and coaching to support home-based exercises. For motivation de five BREQ-2 subscales were converted to a single score by using the relative autonomy index [27]. For the home-based exercise adherence and tablet engagement, after data cleansing an average score was calculated for the 6-month intervention period. These scores were treated as continuous data in further analyses. To compare differences in physical activity level before and after the intervention, a Student paired-samples t test was conducted. The test used a two-sided significance level of .05. Furthermore, multivariate linear regressions were conducted to determine which included variables predict the home-based exercise adherence and the physical activity level. The inclusion of variables as predictors, was based on subject-matter knowledge of the authors. The software package SPSS Statistics version 25 was used to perform the analysis (IBM Corp., 2018).

## 3 Results

### 3.1 Participants

At baseline 133 participants were randomly assigned to an intervention group. The average age was 71.48 (SD 6.39) years old and 92 of the 133 (69.2%) participants were female. In total 103 of the 133 (77.4%) participants completed the 6-month intervention.

### 3.2 Physical Activity Level, Exercise Adherence and Tablet Engagement

At baseline the average physical activity level of participants was 1.49 (SD .14) and after 6 months 1.52 (SD .15). A paired sample t-test revealed the difference between the physical activity level before and after the intervention was not significant, $t_{96}$=-1.63, $P>=.05$.

During the 6-month intervention period, the average app logins per week was 6.72 (SD 4.81). They performed on average 12.53 (SD 11.34) home-based exercises per week and exercised on average 2.88 (SD 1.74) days per week. See Figure 5 to Figure 7 for the weekly progress during the 6-month intervention period.

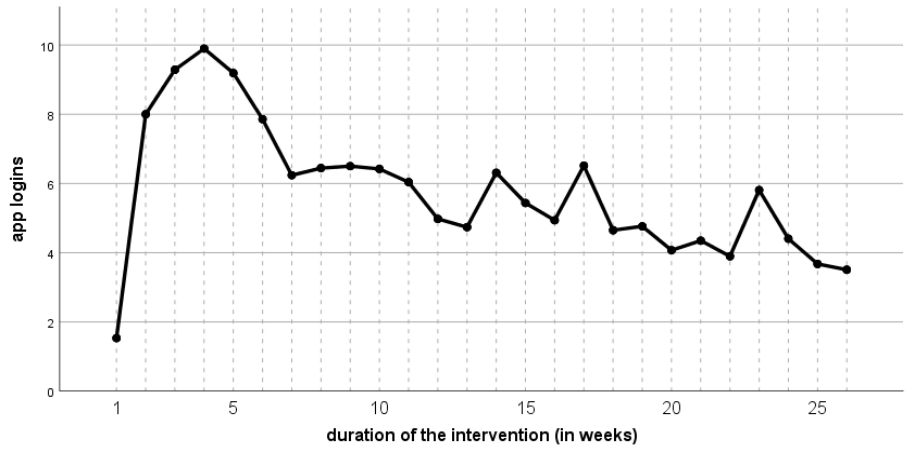

**Fig. 5.** Average number of app logins per week during the 26-week intervention period.

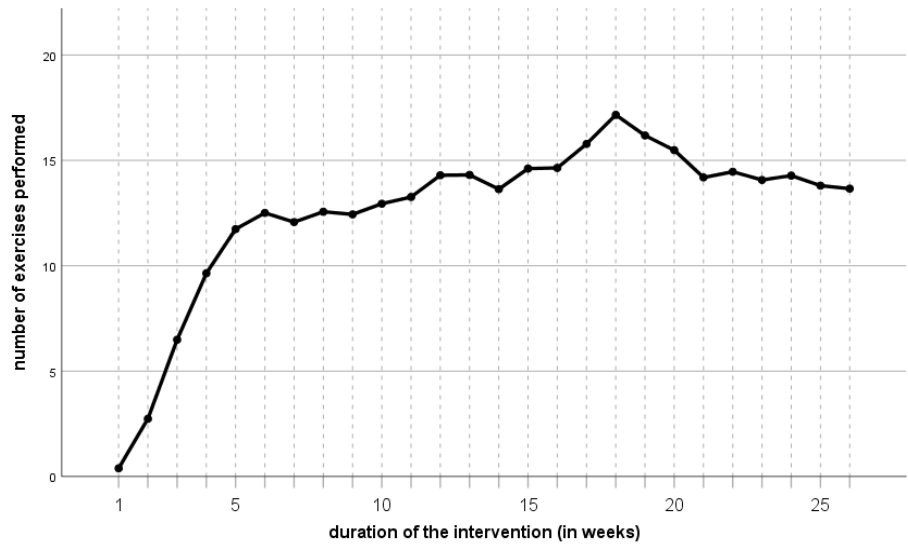

**Fig. 6.** Average number of exercises per week that were performed during the 26-week intervention period.

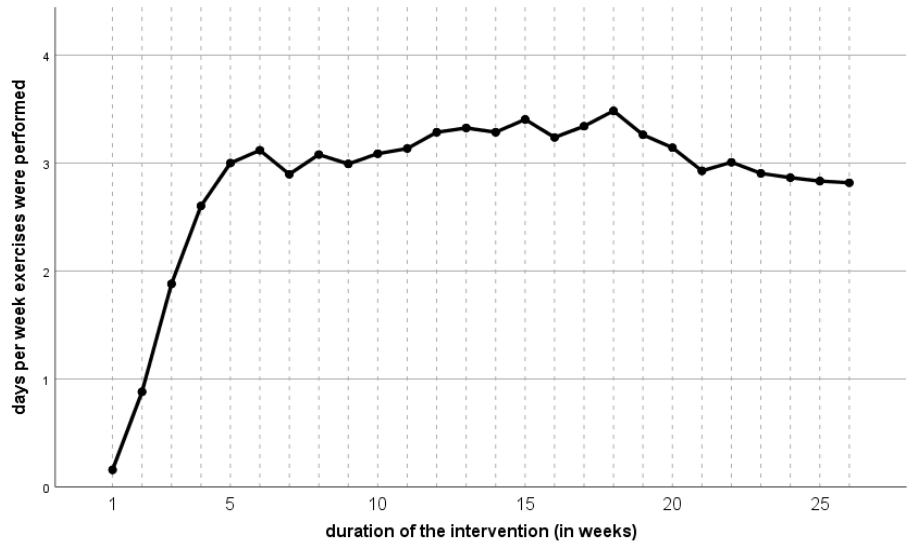

**Fig. 7.** Average number of days per week exercises were performed during the 26-week intervention period.

### 3.3 Prediction of Exercise Adherence

**Prediction of the Number of Exercises Performed Per Week.** A multiple linear regression analysis was performed to predict the number of exercises performed per week, based on age, gender, motivation (baseline), physical activity level (baseline), RCT group, recruitment strategy, season and average app logins. A significant regression equation was found ($F_{8,115}=10.26$, $P<0.001$), with an $R^2$ of .42. Tablet logins, season and recruitment strategy were significant predictors for the number of performed exercises ($P<.0005$, $P=.018$, $P<.0005$, respectively). The number of performed exercises was equal to -2.48 + 1.12 (app logins) + 1.68 (30*offset to midsummer) + 13.43 (recruitment strategy), where recruitment through local community centers is coded as 0, and recruitment through postal mailing is coded as 1. The number of performed exercises increased by 1.12 for each app login, 1.68 for every 30 days the start of the intervention deviated from midsummer and participants recruited through postal mailing performed 13.43 more exercises than participants recruited through local community centers. Age, gender, motivation (baseline), physical activity level (baseline) and RCT group did not significantly predict the number of performed exercises ($P>=.05$).

**Prediction of the Number of Days Per Week Exercises Were Performed.** A second multiple linear regression analysis was performed to predict the number of days per week exercises were performed, based on age, gender, motivation (baseline), physical activity level (baseline), RCT group, recruitment strategy, season and average app logins. A significant regression equation was found ($F_{8,115}=8.81$, $P<0.001$), with an $R^2$ of .38. Tablet logins, season and recruitment strategy were significant predictors for the for number of days per week exercises were performed ($P<.0005$, $P=0.029$, $P=0.007$, respectively). The number of days per week exercises were performed was equal to 1.74 + .20 (app logins) + .24 (30*offset to midsummer) + 1.18 (recruitment strategy), where recruitment through local community centers is coded as 0, and recruitment through postal mailing is coded as 1. The number of days per week exercises were performed increased by .20 for each app login, .24 for every 30 days the start of the intervention deviated from midsummer and participants recruited through postal mailing performed 1.18 more days per week exercises than participants recruited through local community centers. Age, gender, motivation (baseline), physical activity level (baseline) and RCT group did not significantly predict the number of days per week exercises were performed ($P>=.05$).

### 3.4 Prediction of Physical Activity Level

A third multiple linear regression analysis was performed to predict the physical activity level at the end of the intervention (6 months), based on age, gender, motivation (baseline), physical activity level (baseline), RCT group, recruitment strategy, season, average app logins, average number of exercises performed per week and the average number of days per week exercises were performed. A significant regression equation was found ($F_{10,85}=11.74$, $P<0.001$), with an $R^2$ of .58. Age and physical activity level at baseline were significant predictors for physical activity level at 6 months ($P=.043$, $P<.0005$, respectively). The physical activity level at 6 months was equal to .77 – 0.03 (10*age) + .63 (PAL-baseline). The physical activity level at 6 months decreased by 0.03 MET for every 10 years increase of participants' age and increased with .77 MET for each increase of MET at baseline. Gender, motivation (baseline), RCT group, recruitment strategy, season, average app logins, average number of exercises performed per week and the average number of days per week exercises were performed did not significantly predict the physical activity level at 6 months ($P>=.05$).

## 4 Discussion

### 4.1 Principal Findings

The aim of the intervention was to support older adults in performing home-based exercises. In a process evaluation carried out earlier, participants indicated that they felt the tablet and coach were useful [28]. The results of the current study complement these findings. Data derived from the tablet shows that the participants performed on average approximately 13 home-based exercises per week, distributed over 3 days. Participants demonstrated substantive exercise behavior. Participants' engagement with the tablet appears to have contributed to this. The frequency of tablet use predicted the number of days and the number of exercises performed. This affirms the rationale of the intervention that technology can support exercise behavior in older adults.

Although it was not the main goal of the intervention to increase general physical activity of older adults, PA was included as an explorative measure that could provide evidence for secondary effects of the intervention. This appears, however, not to be the case. The participants' physical activity level after completing the 6-month intervention did not differ from their physical activity level before starting the intervention. Also, no association was found between performing home-based exercises and physical activity. Two competing explanations could account for these findings. First, an increased exercise frequency may not lead to a change in the more general notion of physical activity level. Performing a number of exercises per week may not necessarily lead to a more active lifestyle. The sedentary time can remain unchanged, despite the increase in exercise frequency [29]. To stimulate an active lifestyle in older adults, a broader approach is needed [30–32]. A second, competing explanation for the results is that there was an increase of physical activity, but that has gone unnoticed due to the employed procedure of measuring physical activity. The physical activity levels were measured by calculating an average MET-score over a 24-hour period. As a result, a meaningful increase of physical activity for one hour each morning, for instance, may average out over the day. The lack of variation between participants' physical activity levels, indicated by the low SD in the PAL-scores, support this assumption. A solution could be to calculate the duration and average MET only for periods that contain substantive activity, thereby resulting in a more sensitive measure.

Besides intervention characteristics, there were a number of other predictors found. A seasonal effect on exercise behavior was observed. Participants tended to perform more frequent home-based exercises in the winter, compared to the summer. The reason for this could be that older adults prefer outdoor activities, if the weather allows it, thereby limiting the intensity of a home-based exercise program. Physical activity was also predicted by age. Surprisingly, this effect was minute. The strongest predictor was the participants' physical activity at baseline. Existing habits appear to play a dominant role.

### 4.2 Limitations

Previously, a process evaluation was carried out by interviewing participants after they completed the intervention. The conclusions of that study were based on a small number of participants that had to reflect on their behavior over the past 6 months. In contrast, the current study uses log data of the tablet as a more objective and accurate estimations of exercise behavior of all participants. It does not rely on participants' recollection from memory. Nevertheless, also this data source is not impeccable. It cannot be ruled out that participants registered on the tablet that they completed exercises, whilst in reality they did not perform any exercises. The tablet data on exercise completion remains a self-report measure.

Tablet engagement was measured by tracking how often users opened the app. Based on the log data it was, however, not possible to observe how users interacted with the app. Consequently, no conclusions can be drawn which components within the app played a specific role. Furthermore, due to technical issues the number of app logins was not registered flawlessly. As a result, some datapoints on tablet engagement were unusable. To determine if this influenced the results, an alternative measure for tablet activity was used. Participants were asked with a questionnaire how often on average they used the app over the past 6 months. Analysis of this alternative measure was similar to the analysis based on app logins, indicating that use of the app logins was not problematic, despite the technical errors.

This post-hoc study combined data of two groups that participated in the randomized controlled trial; a group that received a tablet and coaching and a group that received tablet, coaching and nutrition advice. Although the two groups differed to which intervention component they were exposed, the analysis used in this study, did not reveal any association between group membership and exercise behavior or physical activity. This validated the choice of the authors to pool the data of both groups, instead of performing the described analysis for both groups separately.

### 4.3 Conclusions

The blended exercise intervention successfully increased the exercise frequency in older adults. Tablet engagement appears to have contributed to this. The frequency of app use predicted the number of exercises and the number of days exercises were performed. The findings suggest that the use of a tablet, in combination with coaching, is a promising strategy to stimulate exercise behavior of older adults. More research is needed how to incorporate general physical activity.